\documentclass{aa}
\usepackage{psfig}
\renewcommand{\(}{\left(}
\renewcommand{\)}{\right)}     
\begin{document}

%
\title{Radial oscillations of relativistic stars}

\author{K. D. Kokkotas \inst{1} \and J. Ruoff\inst{1,2}}

\offprints{K. D. Kokkotas}

\institute{Department of Physics, Aristotle University of
  Thessaloniki,
  Thessaloniki 54006, Greece\\
  email: kokkotas@astro.auth.gr\\
  ruoff@astro.auth.gr\\
  \and Institut f\"ur Astronomie und Astrophysik, Universit\"at
  T\"ubingen, Auf der Morgenstelle 10, 72076 T\"ubingen, Germany }

   
\abstract{ We present a new survey of the radial oscillation modes of
  neutron stars. This study complements and corrects earlier studies
  of radial oscillations. We present an extensive list of frequencies
  for the most common equations of state and some more recent ones.
  In order to check the accuracy, we use two different numerical
  schemes which yield the same results.  The stimulation for this work
  comes from the need of the groups that evolve the full nonlinear
  Einstein equation to have reliable results from perturbation theory
  for comparison.  \keywords{neutron stars -- oscillations of stars --
    equation of state} }

\maketitle


\section{Introduction}
\label{sec:intro}
As they are the simplest oscillation modes of neutron stars, radial
modes have been the first under investigation, more than 35 years ago
(Chandrasekhar 1964a, 1964b). More important, they can give
information about the stability of the stellar model under
consideration. Since radial oscillations do not couple to
gravitational waves, the appropriate equations are quite simple, and
it is relatively easy to numerically solve the eigenvalue problem that
leads to the discrete set of oscillation frequencies of a neutron
star. In the absence of any dissipative processes, the oscillation
spectrum of a stable stellar model forms a complete set; it is
therefore possible to describe any arbitrary periodic radial motion of
a neutron star as a superposition of its various eigenmodes.

The radial modes of neutron stars have been thoroughly investigated by
various authors mostly for zero temperature equations of state (EOS)
(e.g. \cite{Har65}, \cite{Chan77}, \cite{GL83}, \cite{VC92} and
references therein).  But also protoneutron stars with a finite
temperature EOS (\cite{GHZ97}) and strange stars were studied
(\cite{BH91}, \cite{VC92}, \cite{GZ99}).

The first exhaustive compilation of radial modes for various zero
temperature EOS was presented by \cite{GL83} (hereafter GL). However,
as was later pointed out by \cite{VC92} (hereafter VC), their
numerical values for the oscillation frequencies seemed to be flawed
although their equations were correct. VC computed the radial
frequencies for 6 equations of state of dense matter and corroborated
their own results using the argument (\cite{Har65}) that for the
numerical code to be correct, it must yield a zero-frequency mode at
exactly that central density for which the neutron star reaches its
maximal mass. This is the point where the stellar model becomes
unstable with respect to radial collapse if the central density is
further increased. Yet, this is not the case for the results of GL, as
was noticed by VC. However, the above mentioned test can be used only
in the case when both in the stellar model and in the perturbation
equations the equilibrium adiabatic index is used. In general,
different adiabatic indices can be used depending on the physical
conditions inside a star (\cite{GHZ97}). For example, if the slowness
of weak interaction processes are taken into account, the regions of
configurations stable with respect to radial perturbations extend
beyond the central density of the star with the minimum mass
(e.g.~\cite{Chan77}) and of the star with maximum mass (\cite{GHG95}).

In this paper, we repeat the numerical calculation of the radial
oscillation modes of neutron stars for various zero temperature
equations of state using the equilibrium adiabatic index. To verify
the results, we use two different formulations of the equations
together with two different numerical methods to solve the eigenvalue
problem. We find that in all cases we obtain matching values for the
eigenfrequencies. In addition we have verified that the codes yield
zero frequency modes not only at the maxima but also at the minima of
the mass curves.

We give corrected values for the equations of state used by GL, and we
add some new equations of state. It is not clear to us what went wrong
in their calculations, since for certain EOS our values agree with
theirs (EOS C, E, O), for others they differ only slightly (EOS F, L,
N), but for some EOS the discrepancy is quite large (EOS A, B, D, G,
I).

Additionally we include six more recent equations of state: Two models
of \cite{G85}, one of the model of \cite{WFF88}, the EOS MPA of
\cite{Wu91}, and two EOS of \cite{APR98}. Finally, we include three
more tables for polytropic equations of state with different
polytropic indices. The form we use is given by
\begin{equation}
  p = \kappa \rho^{1 + 1/n}\:.
\end{equation}
In particular, we present for the following values of $\kappa$ and
$n$: ($n = 1, \kappa = 100\,$km$^2$), $(n = 0.8, \kappa =
700\,$km$^{2.5}$),$(n = 0.5, \kappa = 2\cdot 10^5\,$km$^4)$.

Another interesting feature is the occurrence of avoided mode
crossings for realistic EOS. This phenomenon has been thoroughly
studied by \cite{GZ99} for a realistic nucleon EOS and an EOS
representing a strange star model. We find that it occurs for all
considered realistic EOS, for some it is quite strongly pronounced,
for others it is less obvious.


\section{Equations and numerical methods}

\subsection{The radial equations}

The static and spherically symmetric metric which describes an
equilibrium stellar model is given by the following line element:
\begin{equation}
  ds^2 = -e^{2\nu} dt^2 + e^{2\lambda} dr^2 + r^2(d\theta^2
  + \sin^2\theta d\phi^2)\;.
\end{equation}
Together with the energy-momentum tensor for a perfect fluid
\begin{equation}
  T_{\mu\nu} = (\rho + p)u_\mu u_\mu + p\,g_{\mu\nu}\;,
\end{equation}
Einstein's field equations yield three independent ordinary
differential equations for the four unknowns $\nu,\mu, \rho$, and $p$.
To complete the set of equations, an equation of state
\begin{equation}
  p = p(\rho)
\end{equation}
must be supplemented. For a given central density, those equations
then yield a unique stellar model with radius $R$ and mass $M$.
Usually one introduces the mass function $m$ via
\begin{equation}
  e^{-2\lambda} = 1 - \frac{2m(r)}{r}
\end{equation}
in order to replace the metric function $\lambda$.

To obtain the equations that govern the radial oscillations, both
fluid and spacetime variables are perturbed in such a way that the
spherical symmetry of the background body is not violated. If we
define as $\delta r(r,t)$ the time dependent radial displacement of a
fluid element located at the position $r$ in the unperturbed model and
assume a harmonic time dependence
\begin{equation}
  \delta r(r,t) = X(r)e^{i\omega t}\;,
\end{equation}
we obtain the following equation describing the radial
oscillations
\begin{eqnarray}
  && C_s^2X''+\left((C_s^2)'-Z+4\pi r \gamma p e^{2\lambda} - \nu'\right)X'\nonumber\\
  &+&\left[2(\nu')^2 + {2m\over r^3}e^{2\lambda} - Z'
    -4\pi(\rho+p)Zre^{2\lambda} +\omega^2 e^{2\lambda-2\nu}\right]X\nonumber\\
  && =0\;,
    \label{master_eqn}
\end{eqnarray}
where $C_s$ is the sound speed, which is calculated from the
unperturbed background for a specific equation of state
\begin{equation}
  C_s^2 = {dp\over d\rho}\;,
\end{equation}
and $\gamma$ is the adiabatic index, which, for adiabatic oscillations,
is related to the sound speed through
\begin{equation}
  \gamma = {\rho +p \over p}{dp \over d\rho}\;.
\end{equation}
Finally
\begin{equation}
  Z(r)=C_s^2\left(\nu' - {2 \over r}\right) \;.
\end{equation}
The boundary condition at the center is that
\begin{equation}
  \delta r (r=0)=0\;,  \label{bc_center}
\end{equation}
while at the surface, the Lagrangian variation of the pressure
should vanish, i.e.
\begin{equation}
  \Delta p =0\;.
\end{equation}
This leads to the condition
\begin{equation}
  \gamma p \zeta(r)' =0\;, \quad \mbox{\rm where}
  \quad \zeta=r^2 e^{-\nu}X\;.
  \label{bc_surface}
\end{equation}
Equation (\ref{master_eqn}) together with the boundary conditions
(\ref{bc_center}) and (\ref{bc_surface}) form a self-adjoint
boundary value problem for $\omega^2$.

As an alternative, the master equation can be written in the variable
$\zeta$ to yield Equation (26.6) of \cite{MTW}, which explicitely
shows its self-adjoint nature:
\begin{equation}\label{MTWev}
        0 = \frac{d}{dr}\(P\frac{d\zeta}{dr}\)
        + \(Q + \omega^2W\)\zeta\;,
\end{equation}
with
\begin{eqnarray}
        r^2 W &=& \(\rho + p\)e^{3\lambda + \nu}\phantom{\bigg(}\\
        r^2 P &=& \gamma p\,e^{\lambda + 3\nu}\\
        r^2 Q &=& e^{\lambda + 3\nu}\(\rho + p\)\((\nu')^2 + 4\frac{\nu'}{r}
        - 8\pi e^{2\lambda}p\)\;.
\end{eqnarray}
At the origin, we have $\zeta(r=0)=0$, and at the surface, the boundary
condition is also given by Eq.~(\ref{bc_surface}).

Since in both the general relativistic and in Newtonian theory, the
oscillation problem is described by a Sturm-Liouville boundary value
problem, the mathematical features that are known for the Newtonian
problem (see \cite{LW58}) also apply to the general relativistic case,
i.e.~the frequency spectrum is discrete, there are $n$ nodes between
the center and the surface of the eigenfunction of the $n$th mode, and
the eigenfunctions are orthogonal.

Since $\omega$ is real for $\omega^2 > 0$, the solution is purely
oscillatory. However for $\omega^2 <0$, the frequency $\omega$ is
imaginary, which corresponds to an exponentially growing solution.
This means that for negative values of $\omega^2$, we have unstable
radial oscillations. For neutron stars, it is the fundamental mode
$\omega_0$ which becomes imaginary at central densities $\rho_c$
larger than the critical density $\rho_{crit}$ for which the total
stellar mass $M$ as a function of $\rho_c$ is maximal. In this case,
the star will ultimately collapse to a black hole. For
$\rho_c=\rho_{crit}$, there frequency of the fundamental mode
$\omega_0$ must vanish. The higher modes become unstable for higher
densities than for maximum mass models.  For realistic equations of
state, there are several regions in the mass-central-density curve,
which are unstable.  On the neutron star branch, there is another
instability point on the low density side, where the star can become
unstable with respect to explosion. This point limits the minimal mass
of a neutron star.

\subsection{The numerical methods}

Since the radial perturbation problem is an old one, various
methods have been used to estimate the radial mode frequencies for
a given equations of state. A numerical integration scheme, which
is similar to what we will describe here, has widely been used,
while a Rayleigh-Ritz variational technique has also been used in
the early times, see \cite{BTM66} for details.

Given the discrepancies existing in the literature, we have
derived the results via two different numerical methods.

The {\em first method} is called in numerical analysis the {\em
shooting method}. In this case, one starts the integration for a
trial value of $\omega^2$ and a given set of initial values of
$X(r=0)$ and $X'(r=0)$ which satisfy at the center the boundary
condition (\ref{bc_center}) and integrates towards the surface.  The
discrete values of $\omega^2$ for which the boundary condition
(\ref{bc_surface}) is satisfied are the eigenfrequencies of the radial
perturbations.

We will apply this method to equation (\ref{MTWev}), but we first
transform it into two first order differential equations.  By
introducing
\begin{equation}
  \eta = P\zeta'\;,
\end{equation}
we obtain
\label{ev_eq}
\begin{eqnarray}
        \label{xi2}\frac{d\zeta}{dr} &=& \frac{\eta}{P}\\
        \frac{d\eta}{dr} &=& -\(\omega^2W + Q\)\zeta\;.
\end{eqnarray}
Through Taylor expansion, we find that close to the origin we have
$\zeta(r) = \zeta_0\,r^3 + {\cal O}(r^5)$ and $\eta(r) = \eta_0 +
{\cal O}(r^2)$. From equation (\ref{xi2}) it then follows that the
leading order coefficients are related by $3\zeta_0 = \eta_0/P(0)$.
Choosing $\eta_0 = 1$, we obtain $\zeta_0 = 1/(3P(0))$, which gives us
the initial values for the integration.

The {\em second method} is based on finite differencing the radial
perturbation equation (\ref{master_eqn}) using second order accurate
schemes for the spatial derivatives. The coefficients of the equation
are calculated for a certain number of, say, $N$ grid points. In this
way a matrix equation of the form
\begin{equation}
  \label{eq:ev}
  \(A - \omega^2_n\mbox{I}\)y = 0\;, \qquad 0 \le n \le N
\end{equation}
is constructed. $A$ is the tridiagonal matrix of the coefficients, I
is the identity matrix, $\omega^2_n$ is the squared frequency of the
$n$th mode, and $y$ is the vector with the unknown values of the
eigenfunction of the specific mode at the $N$ grid points. The
homogeneous linear equation (\ref{eq:ev}) has a nontrivial solution
only if the determinant of the coefficient matrix is equal to zero, i.e.
\begin{equation}
  \label{eq:det}
  \det |A - \omega^2_n \mbox{I}| = 0\;.
\end{equation}
This means that $\omega^2_n$ are the $N$ eigenvalues of the $N\times
N$-matrix $A$. Their numerical evaluation has been achieved using the
routines F01AKF and F02APF of the Nag library.

In this way one can calculate hundreds of radial eigenvalues for a
specific stellar model in a single run. This method is more time
consuming, but one avoids to search for each eigenvalue separately. In
numerical analysis, this method is referred to as {\em Numerov method}.

Using both methods, we have calculated for each stellar model a large
number of eigenvalues, though in the tables of the Appendix, we list
only the three lowest ones. A further check of consistency is that for
each EOS, the maximum mass model must yield zero frequency for the
first mode. This is indeed the case as is it not for the results of
GL.

\section{Results}

Although in principle we could compute the eigenfrequencies up to
arbitrary precision, this would make no sense, since the overall
accuracy of the frequencies is not limited by the machine precision,
but by the number of tabulated values of the equation of state. For
the construction of the stellar background model, one therefore has to
interpolate between the given points. As it turns out, different
interpolation scheme can yield different mode frequencies. Even though
the bulk parameters of the stellar models are not very sensitive to
the actual interpolation scheme, it is the profile of the sound speed,
or equivalently, the adiabatic index which enters into the oscillation
equations, and this quantity is highly sensitive to the interpolation
scheme, especially in the regions where the EOS changes quite
abruptly, such as, for instance, at the neutron drip point.  Since
this region lies in the low pressure regime and is therefore located
close to the surface of the neutron star, it has a quite large
influence on the modes because their amplitudes peak at the surface.
From trying different interpolation schemes, such as linear
logarithmic interpolation or spline interpolation, we find that the
frequencies may vary up to about three per cent. We therefore tabulate
our value with only two significant digits. Only for the polytropic
equations of state, we include three significant digits, since in this
case, the equation of state is analytic, and we do not have to rely on
interpolation.

VC have given tabulated values for EOS D and EOS N. For EOS D, our
results agree with theirs, however, for EOS N we find a
quite significant discrepancy for the stellar parameters like mass and
radius, especially in the high density regime. For instance, for
$\rho_c = 2\times10^{15}\;$g/cm$^3$, they find a mass of $M =
2.563M_\odot$, whereas we obtain $M = 2.621M_\odot$. Since also GL
find the former value, it seems that both GL and VC have used the
tabulated version of the EOS N provided by \cite{LD83}. If we also
use this table, we, again, agree with VC, both
in the stellar parameters and the radial oscillation frequencies.
However, we have access to a table with a larger number of values
(about twice a many in the density range from $10^{14}\;$g/cm$^3$ to
$10^{16}\;$g/cm$^3$), which yields the latter value. In Table \ref{N},
we give the frequencies obtained with the more refined EOS, which,
especially for the fundamental mode, are quite different from the
values of VC. These discrepancies show very drastically that the results
are quite sensitive to the number of tabulated values of a given EOS.

\begin{figure}[h]
  \begin{center}
    \psfig{figure=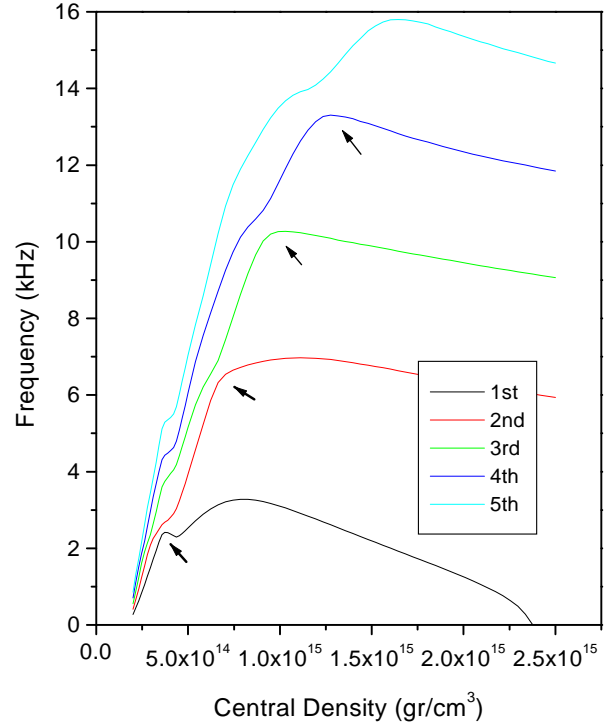,width=8.8cm}
    \caption{We show the first five radial modes as a function of
      the central energy density. The frequency of the fundamental
      mode goes to zero at a density of about $2.35\times
      10^{15}\,$g/cm$^3$, which indicates the onset of radial
      instability with respect to collapse to a black hole. The arrows
      indicate the avoided crossings between the different modes.}
    \label{fig1}
  \end{center}
\end{figure}

In Fig.~1, we show the five radial modes as a function of the central
density for the quite recent EOS APR1 (\cite{APR98}). It is clearly
discernible that the fundamental mode becomes unstable at central
densities above $2.35\times 10^{15}\,$g/cm$^3$. The instability point
corresponds to a stellar model with the maximal allowed mass of $2.38
M_\odot$ and a radius of 10.77\,km.

Another prominent feature is the occurrence of a series of avoided
crossings between the various modes. This peculiarity has been
observed in previous calculations (\cite{GHZ97}), and has been
extensively discussed by \cite{GZ99}. It should be noted that those
avoided crossings do not appear when a polytropic EOS is used (in this
case, one also does not have the second instability point at the low
density region), but it is a characteristic of realistic equations of
state.

The phenomenon of avoided crossings is also known to appear for other
types of oscillations. Depending on the stellar models, there can be
avoided crossings between $g$-modes and $p$-modes. Furthermore,
\cite{AKK96} have reported it to occur between the $f$-mode and the
$w$-modes. Also in rotating stars a similar phenomenon shows up for the
quasi-radial modes when their oscillation frequencies are plotted as a
function of the rotational frequency (\cite{Cl86}, \cite{YE99}).

All these cases have in common that there usually exist two or more
families of modes, which arise from different physical origins.
However, since they are described by a common set of equations, a
particular frequency can only correspond to one single mode.
Therefore, if the frequencies of two modes belonging to different
families start to approach each other, they eventually have to repel
each other before they come too close. This goes along with the modes
exchanging their ``family membership''.

The radial oscillation modes, too, can be divided into two more or
less independent families.  According to \cite{GZ99}, one family lives
predominately in the high density core of the neutron star and the
other in the low density envelope. The two regions are divided by a
``wall'' in the adiabatic index which results from the abrupt change
in the stiffness of the matter at the neutron drip point (c.f. Fig.2
of \cite{GZ99}). This wall effect is present for any realistic EOS,
since it is associated with the neutron drip point, which belongs
to the low pressure regime and is the same for all EOS.

In a model problem, \cite{GZ99} have decoupled both families, and in
this case, both spectra show real crossings, when plotted on top over
each other. When the coupling is brought back, the crossings vanish
and the usual avoided crossing picture reemerges.

\section{Summary}

We have presented updated results for radial oscillations of neutron
stars using a quite exhaustive list of currently available equations
of state, including some very recent ones. For most equations of state
we significantly disagree with the values given by \cite{GL83}. We
have obtained our results by means of two different numerical methods,
which agree up to arbitrary precision.

Furthermore, we have checked that our numerical codes yield a zero
frequency mode located exactly at both instability points, which are
characterized by the local extrema in the mass-density curve. Here, we
also obtain full agreement. The overall accuracy, however, is limited
by the number of tabulated points for a given equation of state. Here,
different numerical interpolation schemes may yield variations in the
frequencies up to about three per cent.

Our results agree with the previous results of \cite{VC92}. However,
we use a more complete table for the EOS N (\cite{S79}), which
significantly alters the values one obtains when the table provided by
\cite{LD83} is used.

We have repeated the calculations for the equations of state already
used by GL, and we have corrected their given values. In
addition, we have included a large number of more recent equations of
state. Since most of the present non-linear evolution codes use
polytropic equations of state, we also have tabulated the mode
frequencies for three different values of the polytropic index $n$.

\begin{acknowledgements}
  J.R. was supported by the Deutsche Forschungsgemeinschaft through
  SFB 382 and the Marie Curie Fellowship No. HPMF-CT-1999-00364.
\end{acknowledgements}

\begin{appendix}
\section{Results for various equations of state}

This appendix provides the numerical data for the radial mode
frequencies of 17 realistic and 3 polytropic EOS. We present the data
in the form of one table for each EOS. In each table we list the
central density $\rho_c$, the radius $R$, and the mass $M$ of the
stellar model, and the frequencies $\nu_n = \omega_n/(2\pi)$ of the
first three radial modes. We also include one stellar model above the
stability limit. For this case, we give the $e$-folding time in $ms$
for the fundamental mode, which is marked by an asterisk.

\begin{table}[h]
\caption{Data for the EOS A (\cite{P71})}
\begin{tabular}{cccccc}
$\rho_c \times 10^{15}$ & $R$ & $M$ & $\nu_0$ & $\nu_1$ &$\nu_2$\\
g/cm$^3$ & km & M$_\odot$ & kHz & kHz & kHz \\
\\
4.200 & 8.335 & 1.654 & 0.34* & 7.55 & 11.91 \\
4.100 & 8.373 & 1.654 & 0.28 & 7.58 & 11.95 \\
3.980 & 8.419 & 1.654 & 0.66 & 7.63 & 12.00 \\
3.000 & 8.874 & 1.621 & 1.97 & 7.98 & 12.33 \\
2.344 & 9.256 & 1.536 & 2.62 & 8.29 & 12.27 \\
1.995 & 9.479 & 1.447 & 2.94 & 8.46 & 11.88 \\
1.698 & 9.667 & 1.329 & 3.23 & 8.57 & 11.31 \\
1.259 & 9.890 & 1.050 & 3.67 & 8.04 & 10.67 \\
\end{tabular}
\end{table}

\begin{table}[h]
\caption{Data for the EOS B (\cite{P71})}
\begin{tabular}{cccccc}
$\rho_c \times 10^{15}$ & $R$ & $M$ & $\nu_0$ & $\nu_1$ &$\nu_2$\\
g/cm$^3$ & km & M$_\odot$ & kHz & kHz & kHz \\
\\
6.100 & 7.024 & 1.413 & 0.32* & 8.75 & 13.18 \\
6.000 & 7.048 & 1.413 & 0.25 & 8.76 & 13.20 \\
5.900 & 7.072 & 1.413 & 0.61 & 8.78 & 13.21 \\
5.500 & 7.175 & 1.411 & 1.30 & 8.86 & 13.32 \\
5.012 & 7.316 & 1.404 & 1.82 & 8.91 & 13.45 \\
3.981 & 7.686 & 1.361 & 2.69 & 8.94 & 13.71 \\
3.388 & 7.953 & 1.304 & 3.10 & 8.86 & 13.71 \\
3.000 & 8.145 & 1.247 & 3.34 & 8.81 & 13.54 \\
1.995 & 8.766 & 0.971 & 3.58 & 8.65 & 11.28 \\
\end{tabular}
\end{table}

\begin{table}
\caption{Data for the EOS C (\cite{BJ74}, model I)}
\begin{tabular}{cccccc}
$\rho_c \times 10^{15}$ & $R$ & $M$ & $\nu_0$ & $\nu_1$ &$\nu_2$\\
g/cm$^3$ & km & M$_\odot$ & kHz & kHz & kHz \\
\\
3.100 & 9.884 & 1.852 & 0.39* 6.23 & 9.55 \\
3.000 & 9.952 & 1.852 & 0.30 & 6.23 & 9.54 \\
2.800 & 10.095 & 1.850 & 0.79 & 6.25 & 9.54 \\
2.500 & 10.326 & 1.840 & 1.24 & 6.28 & 9.54 \\
1.995 & 10.779 & 1.790 & 1.80 & 6.34 & 9.60 \\
1.778 & 11.010 & 1.746 & 2.02 & 6.34 & 9.58 \\
1.413 & 11.443 & 1.619 & 2.36 & 6.33 & 9.47 \\
1.122 & 11.834 & 1.436 & 2.56 & 6.25 & 9.23 \\
1.000 & 12.018 & 1.323 & 2.59 & 6.13 & 8.90 \\
\end{tabular}
\end{table}

\begin{table}
\caption{Data for the EOS D (\cite{BJ74}, model V)}
\begin{tabular}{cccccc}
$\rho_c \times 10^{15}$ & $R$ & $M$ & $\nu_0$ & $\nu_1$ &$\nu_2$\\
g/cm$^3$ & km & M$_\odot$ & kHz & kHz & kHz \\
\\
3.370 & 9.360  & 1.651 & 0.71* & 6.99 & 10.27 \\
3.300 & 9.403  & 1.651 & 0.30 & 6.99 & 10.27 \\
3.000 & 9.598  & 1.648 & 0.84 & 6.96 & 10.32 \\
2.512 & 9.944  & 1.631 & 1.36 & 6.81 & 10.46 \\
1.778 & 10.447 & 1.547 & 2.43 & 6.71 & 9.88  \\
1.413 & 10.678 & 1.424 & 2.96 & 7.25 & 10.37 \\
1.122 & 10.965 & 1.186 & 3.09 & 6.93 & 9.71  \\
\end{tabular}
\end{table}

\begin{table}
\caption{Data for the EOS E (\cite{M74})}
\begin{tabular}{cccccc}
$\rho_c \times 10^{15}$ & $R$ & $M$ & $\nu_0$ & $\nu_1$ &$\nu_2$\\
g/cm$^3$ & km & M$_\odot$ & kHz & kHz & kHz \\
\\
3.000 & 9.061  & 1.726 & 1.78 & 7.62 & 11.56 \\
2.818 & 9.171  & 1.711 & 1.98 & 7.68 & 11.60 \\
2.239 & 9.562  & 1.624 & 2.58 & 7.84 & 11.65 \\
1.778 & 9.915  & 1.474 & 3.02 & 7.90 & 11.44 \\
1.585 & 10.068 & 1.376 & 3.18 & 7.86 & 11.19 \\
1.259 & 10.314 & 1.144 & 3.39 & 7.59 & 10.36 \\
\end{tabular}
\end{table}

\begin{table}
\caption{Data for the EOS F (\cite{A72})}
\begin{tabular}{cccccc}
$\rho_c \times 10^{15}$ & $R$ & $M$ & $\nu_0$ & $\nu_1$ &$\nu_2$\\
g/cm$^3$ & km & M$_\odot$ & kHz & kHz & kHz \\
\\
5.200 & 7.881  & 1.463 & 0.40* & 7.40 & 11.88 \\
5.100 & 7.922  & 1.463 & 0.20 & 7.41 & 11.86 \\
5.012 & 7.961  & 1.463 & 0.46 & 7.42 & 11.83 \\
4.500 & 8.204  & 1.459 & 1.09 & 7.48 & 11.66 \\
3.981 & 8.490  & 1.449 & 1.42 & 7.54 & 11.49 \\
3.162 & 9.088  & 1.412 & 1.63 & 7.40 & 11.15 \\
2.239 & 9.923  & 1.333 & 1.84 & 6.76 & 10.23 \\
1.585 & 10.465 & 1.222 & 2.36 & 6.62 & 9.74  \\
1.122 & 10.889 & 1.032 & 2.75 & 6.57 & 9.03  \\
\end{tabular}
\end{table}

\begin{table}
\caption{Data for the EOS G (\cite{CC74})}
\begin{tabular}{cccccc}
$\rho_c \times 10^{15}$ & $R$ & $M$ & $\nu_0$ & $\nu_1$ &$\nu_2$\\
g/cm$^3$ & km & M$_\odot$ & kHz & kHz & kHz \\
\\
6.300 & 6.945 & 1.357 & 0.70* & 8.77 & 13.54 \\
6.200 & 6.970 & 1.357 & 0.48 & 8.77 & 13.54 \\
6.100 & 6.996 & 1.357 & 0.72 & 8.77 & 13.54 \\
5.800 & 7.075 & 1.356 & 1.18 & 8.79 & 13.53 \\
5.500 & 7.159 & 1.353 & 1.53 & 8.81 & 13.52 \\
5.000 & 7.308 & 1.344 & 2.00 & 8.87 & 13.51 \\
4.503 & 7.472 & 1.327 & 2.40 & 8.95 & 13.53 \\
3.498 & 7.899 & 1.253 & 2.98 & 8.97 & 13.61 \\
2.631 & 8.399 & 1.114 & 3.25 & 8.48 & 12.54 \\
2.376 & 8.557 & 1.057 & 3.35 & 8.34 & 11.86 \\
\end{tabular}
\end{table}

\begin{table}
\caption{Data for the EOS I (\cite{C70})}
\begin{tabular}{cccccc}
$\rho_c \times 10^{15}$ & $R$ & $M$ & $\nu_0$ & $\nu_1$ &$\nu_2$\\
g/cm$^3$ & km & M$_\odot$ & kHz & kHz & kHz \\
\\
2.100 & 11.795 & 2.446 & 0.33* & 5.27 & 8.15 \\
2.000 & 11.900 & 2.447 & 0.24 & 5.32 & 8.23 \\
1.800 & 12.161 & 2.441 & 0.84 & 5.40 & 8.29 \\
1.585 & 12.470 & 2.418 & 1.22 & 5.50 & 8.41 \\
1.259 & 13.025 & 2.324 & 1.69 & 5.60 & 8.52 \\
1.000 & 13.499 & 2.154 & 2.05 & 5.74 & 8.61 \\
0.794 & 13.883 & 1.883 & 2.31 & 5.77 & 8.51 \\
0.631 & 14.127 & 1.561 & 2.46 & 5.67 & 7.95 \\
\end{tabular}
\end{table}

\begin{table}
\caption{Data for the EOS L (\cite{P76})}
\begin{tabular}{cccccc}
$\rho_c \times 10^{15}$ & $R$ & $M$ & $\nu_0$ & $\nu_1$ &$\nu_2$\\
g/cm$^3$ & km & M$_\odot$ & kHz & kHz & kHz \\
\\
1.500 & 13.618 & 2.662 & 0.68* & 4.66 & 7.39 \\
1.400 & 13.747 & 2.660 & 0.56 & 4.70 & 7.47 \\
1.259 & 13.936 & 2.649 & 0.97 & 4.79 & 7.58 \\
1.150 & 14.087 & 2.630 & 1.25 & 4.92 & 7.69 \\
1.000 & 14.299 & 2.579 & 1.59 & 5.21 & 8.05 \\
0.794 & 14.681 & 2.391 & 2.03 & 5.66 & 8.35 \\
0.631 & 14.989 & 2.044 & 2.27 & 5.71 & 8.20 \\
0.600 & 15.025 & 1.959 & 2.32 & 5.69 & 8.15 \\
0.500 & 15.056 & 1.636 & 2.53 & 5.58 & 7.52 \\
0.398 & 14.889 & 1.214 & 2.77 & 5.47 & 6.09 \\
\end{tabular}
\end{table}

\begin{table}
\caption{\label{N}Data for the EOS N (\cite{S79})}
\begin{tabular}{cccccc}
$\rho_c \times 10^{15}$ & $R$ & $M$ & $\nu_0$ & $\nu_1$ &$\nu_2$\\
g/cm$^3$ & km & M$_\odot$ & kHz & kHz & kHz \\
\\
1.700 & 12.740 & 2.634 & 0.47* & 5.10 & 7.97 \\
1.600 & 12.852 & 2.633 & 0.49 & 5.19 & 8.09 \\
1.400 & 13.107 & 2.619 & 1.03 & 5.36 & 8.30 \\
1.200 & 13.385 & 2.575 & 1.47 & 5.62 & 8.57 \\
1.000 & 13.686 & 2.468 & 1.90 & 5.90 & 8.88 \\
0.800 & 13.951 & 2.233 & 2.39 & 6.27 & 9.19 \\
0.600 & 13.980 & 1.729 & 2.96 & 6.61 & 8.79 \\
0.500 & 13.757 & 1.313 & 3.20 & 6.38 & 7.56 \\
0.400 & 13.349 & 0.836 & 3.26 & 5.02 & 5.68 \\
\end{tabular}
\end{table}

\begin{table}
\caption{Data for the EOS O (\cite{BGL75})}
\begin{tabular}{cccccc}
$\rho_c \times 10^{15}$ & $R$ & $M$ & $\nu_0$ & $\nu_1$ &$\nu_2$\\
g/cm$^3$ & km & M$_\odot$ & kHz & kHz & kHz \\
\\
2.100 & 11.502 & 2.379 & 0.55* & 5.56 & 8.64 \\
2.000 & 11.587 & 2.378 & 0.53 & 5.63 & 8.72 \\
1.800 & 11.765 & 2.370 & 1.05 & 5.80 & 8.99 \\
1.600 & 11.974 & 2.346 & 1.43 & 5.94 & 9.23 \\
1.400 & 12.201 & 2.296 & 1.81 & 6.14 & 9.51 \\
1.200 & 12.442 & 2.199 & 2.18 & 6.38 & 9.93 \\
1.000 & 12.672 & 2.019 & 2.56 & 6.75 & 10.21 \\
0.800 & 12.832 & 1.682 & 2.86 & 7.09 & 9.43 \\
0.600 & 12.760 & 1.173 & 3.09 & 6.58 & 7.92 \\
\end{tabular}
\end{table}

\begin{table}
\caption{Data for the EOS G$_{240}$ (\cite{G85})}
\begin{tabular}{cccccc}
$\rho_c \times 10^{15}$ & $R$ & $M$ & $\nu_0$ & $\nu_1$ &$\nu_2$\\
g/cm$^3$ & km & M$_\odot$ & kHz & kHz & kHz \\
\\
2.600 & 10.850 & 1.553 & 0.66* & 5.39 & 8.59 \\
2.500 & 10.928 & 1.553 & 0.33 & 5.40 & 8.58 \\
2.200 & 11.209 & 1.549 & 0.77 & 5.42 & 8.48 \\
1.800 & 11.647 & 1.529 & 1.11 & 5.43 & 8.37 \\
1.400 & 12.200 & 1.481 & 1.37 & 5.34 & 8.10 \\
1.000 & 12.798 & 1.374 & 1.70 & 5.28 & 7.98 \\
0.800 & 13.110 & 1.267 & 1.91 & 5.17 & 7.96 \\
0.600 & 13.351 & 1.095 & 2.27 & 5.27 & 6.86 \\
0.400 & 13.482 & 0.711 & 2.36 & 4.54 & 5.23 \\
\end{tabular}
\end{table}

\begin{table}
\caption{Data for the EOS G$_{300}$ (\cite{G85})}
\begin{tabular}{cccccc}
$\rho_c \times 10^{15}$ & $R$ & $M$ & $\nu_0$ & $\nu_1$ &$\nu_2$\\
g/cm$^3$ & km & M$_\odot$ & kHz & kHz & kHz \\
\\
2.200 & 11.687 & 1.788 & 0.33* & 5.02 & 8.06 \\
2.100 & 11.762 & 1.788 & 0.12 & 5.08 & 8.10 \\
2.000 & 11.842 & 1.787 & 0.51 & 5.15 & 8.15 \\
1.800 & 12.027 & 1.782 & 0.86 & 5.25 & 8.23 \\
1.400 & 12.542 & 1.742 & 1.28 & 5.37 & 8.22 \\
1.000 & 13.182 & 1.624 & 1.66 & 5.36 & 7.99 \\
0.800 & 13.482 & 1.506 & 1.92 & 5.42 & 8.12 \\
0.600 & 13.718 & 1.286 & 2.27 & 5.33 & 7.50 \\
0.500 & 13.733 & 1.119 & 2.54 & 5.61 & 6.60 \\
0.400 & 13.630 & 0.825 & 2.58 & 5.04 & 5.59 \\
\end{tabular}
\end{table}

\begin{table}
\caption{Data for the EOS WFF (\cite{WFF88})}
\begin{tabular}{cccccc}
$\rho_c \times 10^{15}$ & $R$ & $M$ & $\nu_0$ & $\nu_1$ &$\nu_2$\\
g/cm$^3$ & km & M$_\odot$ & kHz & kHz & kHz \\
\\
3.200 & 9.510  & 1.840 & 0.47* & 6.66 & 10.38 \\
3.100 & 9.558  & 1.840 & 0.42 & 6.70 & 10.44 \\
3.000 & 9.612  & 1.840 & 0.69 & 6.73 & 10.45 \\
2.800 & 9.729  & 1.836 & 1.08 & 6.80 & 10.49 \\
2.600 & 9.850  & 1.828 & 1.38 & 6.90 & 10.55 \\
2.000 & 10.278 & 1.759 & 2.13 & 7.20 & 10.89 \\
1.800 & 10.441 & 1.710 & 2.37 & 7.30 & 10.99 \\
1.400 & 10.774 & 1.538 & 2.87 & 7.47 & 11.12 \\
1.200 & 10.925 & 1.389 & 3.12 & 7.52 & 10.78 \\
1.000 & 11.038 & 1.178 & 3.34 & 7.54 & 9.42  \\
0.900 & 11.075 & 1.044 & 3.41 & 7.46 & 8.51  \\
0.800 & 11.104 & 0.889 & 3.42 & 6.99 & 7.68  \\
\end{tabular}
\end{table}

\begin{table}
\caption{Data for the EOS MPA (\cite{Wu91})}
\begin{tabular}{cccccc}
$\rho_c \times 10^{15}$ & $R$ & $M$ & $\nu_0$ & $\nu_1$ &$\nu_2$\\
g/cm$^3$ & km & M$_\odot$ & kHz & kHz & kHz \\
\\
4.800 & 7.899  & 1.560 & 0.39* & 7.82 & 11.96 \\
4.700 & 7.930  & 1.560 & 0.37 & 7.85 & 11.99 \\
4.600 & 7.963  & 1.559 & 0.67 & 7.87 & 12.02 \\
4.500 & 7.996  & 1.559 & 0.87 & 7.90 & 12.06 \\
4.400 & 8.030  & 1.558 & 1.04 & 7.92 & 12.10 \\
4.200 & 8.104  & 1.556 & 1.31 & 7.96 & 12.14 \\
4.000 & 8.186  & 1.551 & 1.55 & 7.99 & 12.16 \\
3.500 & 8.407  & 1.531 & 2.07 & 8.08 & 12.26 \\
3.000 & 8.669  & 1.489 & 2.51 & 8.13 & 12.24 \\
2.500 & 8.973  & 1.410 & 2.90 & 8.16 & 12.10 \\
2.000 & 9.328  & 1.269 & 3.19 & 8.08 & 11.61 \\
1.500 & 9.747  & 1.033 & 3.29 & 7.58 & 10.48 \\
1.200 & 10.031 & 0.844 & 3.27 & 6.95 & 9.00 \\
1.000 & 10.251 & 0.698 & 3.22 & 6.36 & 7.47 \\
\end{tabular}
\end{table}

\clearpage

\begin{table}
\caption{Data for the EOS APR1 (\cite{APR98})}
\begin{tabular}{cccccc}
$\rho_c \times 10^{15}$ & $R$ & $M$ & $\nu_0$ & $\nu_1$ &$\nu_2$\\
g/cm$^3$ & km & M$_\odot$ & kHz & kHz & kHz \\
\\
2.400 & 10.746 & 2.379 & 0.40* & 6.01 & 9.14 \\
2.300 & 10.822 & 2.379 & 0.47 & 6.08 & 9.21 \\
2.200 & 10.904 & 2.377 & 0.80 & 6.16 & 9.29 \\
2.100 & 10.990 & 2.373 & 1.04 & 6.24 & 9.37 \\
2.000 & 11.080 & 2.366 & 1.25 & 6.32 & 9.45 \\
1.800 & 11.277 & 2.340 & 1.64 & 6.50 & 9.62 \\
1.500 & 11.611 & 2.250 & 2.19 & 6.76 & 9.89 \\
1.200 & 11.966 & 2.040 & 2.75 & 6.96 & 10.16 \\
1.000 & 12.171 & 1.774 & 3.10 & 6.94 & 10.27 \\
0.800 & 12.294 & 1.365 & 3.28 & 6.75 & 8.82 \\
0.700 & 12.336 & 1.109 & 3.21 & 6.53 & 7.39 \\
0.600 & 12.435 & 0.841 & 2.95 & 5.52 & 6.37 \\
\end{tabular}
\end{table}

\begin{table}
\caption{Data for the EOS APR2 (\cite{APR98})}
\begin{tabular}{cccccc}
$\rho_c \times 10^{15}$ & $R$ & $M$ & $\nu_0$ & $\nu_1$ &$\nu_2$\\
g/cm$^3$ & km & M$_\odot$ & kHz & kHz & kHz \\
\\
2.800 & 9.998 & 2.201 & 0.39* & 6.43 & 9.71 \\
2.700 & 10.059 & 2.201 & 0.45 & 6.50 & 9.79 \\
2.600 & 10.122 & 2.199 & 0.77 & 6.57 & 9.88 \\
2.500 & 10.193 & 2.197 & 1.01 & 6.63 & 9.92 \\
2.400 & 10.269 & 2.192 & 1.21 & 6.69 & 9.98 \\
2.200 & 10.428 & 2.176 & 1.58 & 6.83 & 10.08 \\
2.000 & 10.598 & 2.148 & 1.92 & 6.98 & 10.22 \\
1.800 & 10.789 & 2.098 & 2.25 & 7.12 & 10.35 \\
1.400 & 11.203 & 1.890 & 2.89 & 7.26 & 10.54 \\
1.000 & 11.572 & 1.410 & 3.37 & 7.01 & 10.07 \\
0.800 & 11.737 & 1.032 & 3.25 & 6.59 & 7.58 \\
0.700 & 11.884 & 0.826 & 3.01 & 5.88 & 6.54 \\
0.600 & 12.189 & 0.632 & 2.63 & 4.50 & 5.76 \\
\end{tabular}
\end{table}

\begin{table}
\caption{Data for the polytropic EOS with $n=1$ and $\kappa =
100\,$km$^2$}
\begin{tabular}{cccccc}
$\rho_c \times 10^{15}$ & $R$ & $M$ & $\nu_0$ & $\nu_1$ &$\nu_2$\\
g/cm$^3$ & km & M$_\odot$ & kHz & kHz & kHz \\
\\
5.700 & 7.518 & 1.351 & 0.618* & 7.582 & 11.569 \\
5.650 & 7.535 & 1.351 & 0.180 & 7.576 & 11.556 \\
5.600 & 7.554 & 1.351 & 0.358 & 7.569 & 11.542 \\
5.500 & 7.590 & 1.351 & 0.569 & 7.557 & 11.520 \\
5.300 & 7.667 & 1.350 & 0.838 & 7.524 & 11.457 \\
5.000 & 7.787 & 1.348 & 1.129 & 7.475 & 11.365 \\
4.000 & 8.256 & 1.326 & 1.755 & 7.244 & 10.950 \\
3.000 & 8.862 & 1.266 & 2.141 & 6.871 & 10.319 \\
2.000 & 9.673 & 1.126 & 2.323 & 6.237 & 9.295 \\
1.500 & 10.19 & 0.998 & 2.302 & 5.737 & 8.513 \\
1.000 & 10.81 & 0.802 & 2.150 & 5.007 & 7.394 \\
\end{tabular}
\end{table} 

\begin{table}
\caption{Data for the polytropic EOS with $n=0.8$ and $\kappa =
700\,$km$^{2.5}$}
\begin{tabular}{cccccc}
$\rho_c \times 10^{15}$ & $R$ & $M$ & $\nu_0$ & $\nu_1$ &$\nu_2$\\
g/cm$^3$ & km & M$_\odot$ & kHz & kHz & kHz \\
\\

4.800 & 7.832 & 1.609 & 0.630* & 7.601 & 11.626 \\
4.750 & 7.853 & 1.609 & 0.281 & 7.602 & 11.624 \\
4.700 & 7.874 & 1.609 & 0.459 & 7.604 & 11.623 \\
4.600 & 7.917 & 1.609 & 0.690 & 7.606 & 11.618 \\
4.500 & 7.961 & 1.608 & 0.862 & 7.608 & 11.612 \\
4.300 & 8.053 & 1.606 & 1.132 & 7.609 & 11.596 \\
4.000 & 8.199 & 1.600 & 1.455 & 7.610 & 11.573 \\
3.500 & 8.470 & 1.579 & 1.868 & 7.577 & 11.477 \\
3.000 & 8.778 & 1.539 & 2.199 & 7.501 & 11.314 \\
2.500 & 9.126 & 1.468 & 2.464 & 7.361 & 11.051 \\
2.000 & 9.509 & 1.351 & 2.656 & 7.121 & 10.636 \\
1.500 & 9.908 & 1.161 & 2.741 & 6.711 & 9.969 \\
1.000 & 10.251 & 0.865 & 2.649 & 5.998 & 8.856 \\
\end{tabular}
\end{table} 

\begin{table}
\caption{Data for the polytropic EOS with $n=0.5$ and 
$\kappa = 2\cdot 10^5\,$km$^4$}
\begin{tabular}{cccccc}
$\rho_c \times 10^{15}$ & $R$ & $M$ & $\nu_0$ & $\nu_1$ &$\nu_2$\\
g/cm$^3$ & km & M$_\odot$ & kHz & kHz & kHz \\
\\
3.500 & 8.604 & 2.120 & 0.322* & 7.308 & 11.220 \\
3.450 & 8.629 & 2.120 & 0.244 & 7.344 & 11.268 \\
3.400 & 8.655 & 2.120 & 0.620 & 7.402 & 11.347 \\
3.300 & 8.708 & 2.120 & 0.848 & 7.452 & 11.412 \\
3.200 & 8.763 & 2.118 & 1.140 & 7.548 & 11.542 \\
3.000 & 8.881 & 2.111 & 1.519 & 7.698 & 11.740 \\
2.600 & 9.140 & 2.075 & 2.151 & 8.005 & 12.139 \\
2.200 & 9.419 & 1.988 & 2.716 & 8.305 & 12.516 \\
1.800 & 9.672 & 1.809 & 3.235 & 8.555 & 12.804 \\
1.400 & 9.784 & 1.484 & 3.665 & 8.651 & 12.850 \\
1.200 & 9.713 & 1.252 & 3.792 & 8.551 & 12.653 \\
1.000 & 9.491 & 0.977 & 3.870 & 8.369 & 12.334 \\
0.800 & 9.045 & 0.678 & 3.810 & 7.962 & 11.690 \\
\end{tabular}
\end{table} 

\end{appendix}

\end{document}